\documentstyle[12pt,psfig]{article}
\makeatletter
\def\@cite#1#2{{[{#1}]\if@tempswa\typeout
{IJCGA warning: optional citation argument
ignored: `#2'} \fi}}

\newcount\@tempcntc
\def\@citex[#1]#2{\if@filesw\immediate\write\@auxout{\string\citation{#2}}\fi
  \@tempcnta\z@\@tempcntb\m@ne\def\@citea{}\@cite{\@for\@citeb:=#2\do
    {\@ifundefined
       {b@\@citeb}{\@citeo\@tempcntb\m@ne\@citea\def\@citea{,}{\bf ?}\@warning
       {Citation `\@citeb' on page \thepage \space undefined}}%
    {\setbox\z@\hbox{\global\@tempcntc0\csname b@\@citeb\endcsname\relax}%
     \ifnum\@tempcntc=\z@ \@citeo\@tempcntb\m@ne
       \@citea\def\@citea{,}\hbox{\csname b@\@citeb\endcsname}%
     \else
      \advance\@tempcntb\@ne
      \ifnum\@tempcntb=\@tempcntc
      \else\advance\@tempcntb\m@ne\@citeo
      \@tempcnta\@tempcntc\@tempcntb\@tempcntc\fi\fi}}\@citeo}{#1}}
\def\@citeo{\ifnum\@tempcnta>\@tempcntb\else\@citea\def\@citea{,}%
  \ifnum\@tempcnta=\@tempcntb\the\@tempcnta\else
   {\advance\@tempcnta\@ne\ifnum\@tempcnta=\@tempcntb \else
\def\@citea{--}\fi
    \advance\@tempcnta\m@ne\the\@tempcnta\@citea\the\@tempcntb}\fi\fi}
\makeatother

\def\psla{p\kern-.45em/}
\def\qsla{q\kern-.45em/}
\def\ksla{k\kern-.45em/}
\def\dels{\partial\kern-.45em/}

\newcommand{\dr}{\mbox{\footnotesize$\overline{\rm DR}{}$}}
\newcommand{\ms}{\mbox{\footnotesize$\overline{\rm MS}{}$}}

\newcommand{\tilt}{\tilde{t}}

\newcommand{\tilb}{\tilde{b}}

\newcommand{\gsim}{\lower.7ex\hbox{$\;\stackrel{\textstyle>}{\sim}\;$}}
\newcommand{\lsim}{\lower.7ex\hbox{$\;\stackrel{\textstyle<}{\sim}\;$}}
\newcommand{\be}{\begin{equation}}
\newcommand{\ee}{\end{equation}}
\newcommand{\bea}{\begin{eqnarray}}
\newcommand{\eea}{\end{eqnarray}}

\def\baselinestretch{1}
\begin{document}
\catcode`@=11
\newtoks\@stequation
\def\subequations{\refstepcounter{equation}%
\edef\@savedequation{\the\c@equation}%
  \@stequation=\expandafter{\theequation}
  \edef\@savedtheequation{\the\@stequation}
  \edef\oldtheequation{\theequation}%
  \setcounter{equation}{0}%
  \def\theequation{\oldtheequation\alph{equation}}}
\def\endsubequations{\setcounter{equation}{\@savedequation}%
  \@stequation=\expandafter{\@savedtheequation}%
  \edef\theequation{\the\@stequation}\global\@ignoretrue

\noindent}
\catcode`@=12
\begin{titlepage}

\title{{\bf  
Scale- and gauge-independent mixing angles
for scalar particles}}
\vskip2in
\author{  
{\bf J.R. Espinosa$^{1,2}$\footnote{\baselineskip=16pt E-mail address: {\tt
espinosa@makoki.iem.csic.es}}} and 
{\bf Y. Yamada$^{3}$\footnote{\baselineskip=16pt E-mail address: {\tt
yamada@tuhep.phys.tohoku.ac.jp}}}
\hspace{3cm}\\
 $^{1}$~{\small I.M.A.F.F. (CSIC), Serrano 113 bis, 28006 Madrid, Spain}
\hspace{0.3cm}\\
 $^{2}$~{\small I.F.T. C-XVI, U.A.M., 28049 Madrid, Spain}
\hspace{0.3cm}\\
 $^{3}$~{\small Department of Physics, Tohoku University, 
Sendai 980-8578, Japan}
} 
\date{} 
\maketitle 
\def\baselinestretch{1.15} 
\begin{abstract}
\noindent 
The existing definitions of mixing angles (one-loop radiatively
corrected and renormalization-scale independent) for scalar particles
turn out to be gauge dependent when used in gauge theories. We show
that a scale- and gauge-independent mixing angle can be obtained if the
scalar self-energy is improved by the pinch technique, and give two
relevant examples in the Minimal Supersymmetric Standard Model: the
mixing of CP-even Higgs scalars and of top squarks. We also show that the
recently proposed definition of mixing angle that uses the (unpinched)
scalar two-point function evaluated at a particular value of the
external momentum [$p_*^2=(M_1^2+M_2^2)/2$, where $M_{1,2}$ are the
masses of the mixed particles] computed in the Feynman gauge coincides with
the gauge-invariant pinched result. In alternative definitions ({\it e.g.} 
in the on-shell scheme), the improved Higgs mixing angle is 
different from that in the Feynman gauge. Some freedom in the pinch 
technique for scalar-scalar-gauge couplings is also discussed. 
\end{abstract}

\thispagestyle{empty}
\vspace{5cm}
\leftline{July 2002}
\leftline{}

\vskip-24cm
\rightline{}
\rightline{IFT-UAM/CSIC-02-33}
\rightline{TU-665}
\rightline{hep-ph/0207351}
\vskip3in

\end{titlepage}
\setcounter{footnote}{0} \setcounter{page}{1}
\newpage
\baselineskip=20pt

\section{Introduction}

The mixing among fermions with the same quantum numbers is a very important
aspect of the Standard Model (in the quark and neutrino sectors). 
In this paper we focus on the mixing of scalar particles that will 
play a similarly important role if the world is supersymmetric at low energy. 
In particular, the Minimal Supersymmetric Standard Model 
(MSSM) \cite{MSSM} accommodates two main examples of
interest: the mixing among Higgs bosons (of the two CP-even scalars if CP is
conserved; of three states if CP violation in the Higgs sector is important) 
and the mixing among squarks (with the case of top squarks being the 
one in which such effects are expected to be larger). 
Both cases are sensitive to the large top Yukawa coupling and 
radiative corrections to the mixing matrices are significant.

There have been several proposals for the renormalized mixing matrix 
of scalar bosons, which both are independent of the renormalization scale 
[unlike what happens in the modified minimal subtraction ($\ms$) scheme], and 
do not rely on the specific process considered. One is the on-shell scheme
\cite{GuaschEberl}. Another, which we call the $p_*$ scheme, was recently
proposed in Ref.~\cite{EN}. In both
schemes (discussed in more detail in section~2) 
the counterterm for the mixing matrix is constructed from the 
off-diagonal self-energies $\Pi_{ij}(p^2)$ of the scalar bosons. 
Unfortunately, the on-shell scheme was shown to be dependent on the 
gauge fixing parameters $\xi$ in general $R_{\xi}$ gauges \cite{Yamada}
(the same happens with the on-shell fermion mixing matrices
\cite{CKM,Pila}). 
This is also the case for the latter scheme \cite{EN} 
as we show in section~2. 

One way of avoiding this difficulty is to apply a procedure to build
gauge-independent self-energies, so that the counterterm 
of the mixing matrix given in terms of these improved
self-energies is automatically gauge independent. We perform this
improvement by using the well known pinch technique \cite{pinchsym,PaPi,Pa}.
In section~3 we carry out this program for the top squark sector, 
while in section~4 we do the same for the (CP-conserving) Higgs sector, 
after showing explicitly the
gauge dependence of the Higgs boson self-energies in the $R_\xi$ gauge.
In both cases we arrive at an improved definition (in either the on-shell 
or the $p_*$
scheme) of mixing matrices (or, equivalently, mixing angles) that is not only
scale and process independent but also gauge independent. 
There is, however, some freedom regarding the inclusion of 
some pinched contributions in the improved self-energies. 
This problem is addressed, by using the background field
method, in section~5 where we arrive at a prescription to determine what
contributions to include in the pinched self-energies.
A brief summary and our conclusions regarding the prescription
to obtain gauge-independent mixing angles are presented in section~6.

\section{Scale-independent renormalization schemes for \\ the 
mixing matrix of scalars}

In this section we review two schemes for the renormalization of the
mixing matrices of scalar particles that are independent of the
renormalization scale $Q$, and show that both of them are generally
dependent on the gauge fixing parameters.  We first consider the running
of the mixing matrix of $n$ real scalar particles, in the $\ms$ [or 
dimensional reduction $\dr$ for supersymmetric theories] 
renormalization scheme. It is assumed that
these particles do not mix with massive gauge bosons. The generalization
for complex scalars is straightforward.

Let $\phi_{\alpha}$ ($\alpha=1,\ldots, n$) be real scalar fields 
in the gauge eigenstate basis, related to the bare fields $\phi_{\alpha 
0}$ by
\be
\phi_{\alpha 0} = \left( \delta_{\alpha\beta} 
+ \frac{1}{2}\delta Z_{\alpha\beta} \right) \phi_{\beta}\ .
\ee
(Unless otherwise indicated, a sum over repeated
indices is always understood.)
The running mass matrix $M^2_{\alpha\beta}$, related to bare parameters
and counterterms by
\be
M^2_{\alpha\beta}=M^2_{\alpha\beta 0}-\delta M^2_{\alpha\beta}
+{1\over 2}(\delta
Z_{\gamma\alpha}M_{\gamma\beta}^2+M_{\alpha\gamma}^2\delta 
Z_{\gamma\beta})\ ,
\ee 
is diagonalized by a 
real orthogonal mixing matrix $R$ as 
\be
m_i^2\delta_{ij} = R_{i\alpha} M^2_{\alpha\beta} R_{j\beta}
 \label{eqrunR}
\ee
(with no sum over $i$).
The relation between the gauge eigenstates $\phi_{\alpha}$ and the
mass eigenstates $\phi_i$, with (running) masses $m_i$, is 
then expressed as 
\be
\phi_i= R_{i\alpha}\phi_{\alpha}\ ,\;\;\;
\phi_{\alpha}= R_{i\alpha}\phi_i\ ,
\ee
with $R_{i\alpha}R_{i\beta}=\delta_{\alpha\beta}$ and 
$R_{i\alpha}R_{j\alpha}=\delta_{ij}$.

The $Q$ dependence of the running mixing matrix $R_{i\alpha}$ is 
determined from the $i\neq j$ parts of Eq.~(\ref{eqrunR}). 
Using
\bea
\label{rge1}
\frac{d \phi_{\alpha}}{d\ln Q^2}  
& = &  \gamma_{\alpha\beta}\phi_\beta\ , \\
\frac{d M^2_{\alpha\beta}}{d\ln Q^2} &=& \beta^0_{\alpha\beta} 
- M^2_{\alpha\rho}\gamma_{\rho\beta} 
- \gamma_{\rho\alpha} M^2_{\rho\beta}\ ,
\label{rge2}
\eea
where $\beta^0_{\alpha\beta}=-d \delta M^2_{\alpha\beta}/d\ln Q^2$,
we obtain 
\bea
\label{dRQ}
\frac{d}{d\ln Q^2} R_{i\alpha} &=& \sum_{j\neq i}X_{ij}R_{j\alpha}\ , \\
X_{ij} &=& \frac{1}{m_i^2-m_j^2}[ \beta^0_{ij} - (m_i^2+m_j^2)\gamma_{ij} ]\ .
\label{Xij}
\eea
In this expression $\beta^0_{ij}$ and $\gamma_{ij}$ are obtained from 
$\beta^0_{\alpha\beta}$ and $\gamma_{\alpha\beta}$, respectively, 
by rotating with $R$. For $R$ to be orthogonal at any $Q$, $X_{ij}$ must 
satisfy $X_{ij}+X_{ji}=0$, as is realized in Eq.~(\ref{Xij}). 

Although the $\ms$ ($\dr$) running mixing matrix is simple, it is often
convenient to introduce an effective mixing matrix that is independent of
the renormalization scale $Q$. In addition, to be independent of specific
processes that involve $\phi_i$ fields, it is desirable to construct the
counterterm $\delta R$ using the self-energy $\Pi_{ij}(p^2)$, after
discarding the absorptive part. 
Here we show two examples of such schemes
for the case of two mixed particles ($n=2$; in this case, $R_{i\alpha}$ is
parametrized by one number, the mixing angle).  The first example is the
on-shell scheme \cite{GuaschEberl}, where $\delta R$ is fixed to absorb the
anti-hermitian part of the wave function correction $\delta Z_{ij}$ for an
external on-shell particle $\phi_j$.  Its relation to the running $R$ is
\be
R^{OS}_{i\alpha} = R_{i\alpha} 
- \sum_{j\neq i} 
\frac{\Pi_{ij}(m_i^2)+\Pi_{ij}(m_j^2)}{2(m_i^2-m_j^2)}R_{j\alpha}\ .
\label{ROS}
\ee
The bare one-loop two-point
functions $\Pi_{\alpha\beta 0}(p^2)$ are in general ultraviolet (UV) divergent.
In the $\ms$ ($\dr$) scheme, this divergence leads to a dependence on
the renormalization scale $Q$ of the renormalized two-point function
$\Pi_{\alpha\beta}(p^2)$ given by 
\be
\frac{\partial}{\partial \ln Q^2} \Pi_{\alpha\beta}(p^2) = 
\beta_{\alpha\beta}^0 -  p^2
(\gamma_{\alpha\beta}+\gamma_{\beta\alpha}) \, . 
\label{dPiQ}
\ee
To show this, relate the renormalized one-loop inverse propagator
$\Gamma_{\alpha\beta}(p^2)$ 
to the bare inverse propagator $\Gamma_{\alpha\beta 0}(p^2)$ to get
\be
\Gamma_{\alpha\beta 0}(p^2)= p^2\delta_{\alpha\beta} - M^2_{\alpha\beta} 
+ \Pi_{\alpha\beta}(p^2) 
-{1\over 2} p^2 (\delta Z_{\alpha\beta}+\delta Z_{\beta\alpha}) 
+{1\over 2}( M^2_{\alpha\gamma}\delta Z_{\gamma\beta} 
+ M^2_{\gamma\beta}\delta Z_{\gamma\alpha})\ .
\ee
{}From the 
condition that the bare inverse propagator $\Gamma_{\alpha\beta 0}(p^2)$ 
is $Q$ independent, and using Eqs.~(\ref{rge1}) and (\ref{rge2}),
Eq.~(\ref{dPiQ}) follows.
Now, using Eqs.~(\ref{dPiQ}) and (\ref{dRQ}), $R^{OS}$ is shown to be 
$Q$ independent.

More recently, Ref.~\cite{EN} proposed another definition of the
mixing angle as the one that appears in the rotation that diagonalizes the
effective mass matrix (for $n=2$), 
\be
\overline{M}^2_{\alpha\beta}(p^2) \equiv M^2_{\alpha\beta} 
-\Pi_{\alpha\beta}(p^2). 
\ee
This mixing matrix is 
\be
\label{Rp}
[R(p^2)]_{i\alpha} = R_{i\alpha} 
- \sum_{j\neq i}\frac{\Pi_{ij}(p^2)}{m_i^2-m_j^2}R_{j\alpha}\ .
\ee
Using Eqs.~(\ref{dPiQ}) and (\ref{dRQ}), it is seen that
$[R(p^2)]$ 
becomes $Q$ independent at $p^2=p^2_*\equiv(m_1^2+m_2^2)/2$. A 
scale-independent renormalized mixing angle ($p_*$ scheme) is 
then given by the elements of $[R(p_*^2)]_{i\alpha}$. 

In gauge theories, individual vertex functions generally 
depend on the gauge fixing, while the total corrections to physical 
quantities (masses, cross sections, etc.) and the gauge-symmetric 
$\ms$ ($\dr$) parameters do not. It is therefore 
necessary to examine the gauge dependence of the scale-independent 
mixing angles defined above. Working in the $R_{\xi}$ gauge,
the
dependence of $\Pi_{ij}(p^2)$ on the gauge parameter $\xi$ is given by 
the Nielsen identity \cite{nielsen,Gambino}\ , 
\be
\partial_{\xi} \Pi_{ij}(p^2) = 
\Lambda_{ij}(p^2) (p^2 - m_j^2) 
+ (p^2 - m_i^2) \Lambda^*_{ji}(p^2)\ , 
\label{eq22}
\ee
where $\Lambda_{ij}(p^2)$ are some one-loop scalar functions. This
identity is crucial to show that pole masses are gauge independent and
can be used to study also the gauge dependence of mixing angles. In both
the on-shell and the $p_*$ schemes, the $\xi$ 
dependence remains in $\delta R$, implying the gauge dependence of these 
schemes. More explicitly, one gets
\be
\partial_\xi R^{OS}_{i\alpha}=-{1\over 2}\sum_{j\neq i}R_{j\alpha}
\left[\Lambda_{ij}(m_i^2)-\Lambda_{ji}^*(m_j^2)
\right]\neq 0
\ee
for the on-shell mixing matrix and
\be
\partial_\xi [R(p^2)]_{i\alpha}=-{1\over 2}\sum_{j\neq i}R_{j\alpha}
\left[\Lambda_{ij}(p_*^2)-\Lambda_{ji}^*(p_*^2)
\right]\neq 0
\ee
for the $p_*$ scheme (in general, $\Lambda$ is not a hermitian matrix).
The gauge-dependent parts of $\delta R$ are UV 
finite \cite{Yamada,Pila} and numerically rather small, but not 
satisfactory from the theoretical point of view. 

Nevertheless, we may modify these scale-independent mixing angles to be 
gauge independent. This is done by splitting the gauge-dependent parts 
from the off-diagonal self-energies $\Pi_{ij}(p^2)$ by 
a definite procedure, and then using the remaining, 
gauge-independent parts for the definitions (\ref{ROS}) and (\ref{Rp}). 
In the following two sections, we perform such a modification for two 
cases in the MSSM, adopting the pinch technique: the left-right mixing 
of top squarks and the mixing of two CP-even Higgs bosons. 

\section{Mixing angle of top squarks}

The gauge eigenstates $\tilt_{\alpha}=(\tilt_L, \tilt_R)$ 
of top squarks mix with each other to give the mass 
eigenstates $\tilt_i$ $(i=1,2)$. Their relation is  
given by 
$\tilt_i=R^{\tilt}_{i\alpha}\tilt_{\alpha}$ with the left-right mixing matrix 
\be 
R^{\tilt}_{i\alpha}=\left(
\begin{array}{cc}\cos\theta_{\tilt} & \sin\theta_{\tilt} \\
-\sin\theta_{\tilt} & \cos\theta_{\tilt} \end{array}\right)\ . 
\label{eq34}
\ee
We first consider the gauge dependence of the scale-independent
renormalization of the left-right mixing angle of the top squarks
$\tilde{t}$ in general $R_{\xi}$ gauges, and later on discuss its
improvement by the pinch technique.

The unrenormalized two-point function $\Pi^{\tilt}_{ij}(p^2)$ 
is gauge dependent. Its dependence on $\xi_Z$ and $\xi_W$,
expressed as the difference from the Feynman gauge result \cite{pierce}, 
takes the following form \cite{Yamada}
\bea
\Pi^{\tilt}_{ij}(p^2)&=  
&\left.\Pi^{\tilt}_{ij}(p^2)\right|_{\xi_Z=\xi_W=1}\nonumber\\ 
&+&\frac{g_Z^2}{16\pi^2}(1-\xi_Z)
\chi^Z_{ik}\chi^Z_{jk}\left[ 
-f_{ij}(p^2)\alpha_Z +g_{ij}(p^2,m_{\tilt_k}^2)
\beta^{(0)}_{Z\tilt_k}(p^2) \right] \nonumber\\
&+&\frac{g_2^2}{8\pi^2}(1-\xi_W)
\chi^W_{ik}\chi^W_{jk} \left[
-f_{ij}(p^2)\alpha_W +g_{ij}(p^2,m_{\tilb_k}^2)
\beta^{(0)}_{W\tilb_k}(p^2)  \right]\, . \label{eq36}
\eea
Here $g_Z^2=g_1^2+g_2^2$, with $g_1$ and $g_2$ the gauge coupling
constants of $U(1)_Y$ and $SU(2)_L$, respectively, and we have defined the
quantities ($R^{\tilb}_{i\alpha}$ is the mixing matrix of bottom squarks)
\bea
\chi^Z_{ik}&\equiv &
{1\over 2}R^{\tilt}_{iL}R^{\tilt}_{kL}-{2\over 3}\delta_{ik}\sin^2\theta_W\ ,\\
\chi^W_{ik}&\equiv & {1\over 2}R^{\tilt}_{iL}R^{\tilb}_{kL}\ ,
\eea
the functions
\bea
\label{f}
f_{ij}(p^2)&\equiv & {1\over 2} (2p^2-m_i^2-m_j^2)\ ,\\
g_{ij}(p^2,m^2)&\equiv &
(p^2-m^2)(2p^2-m_i^2-m_j^2)-(p^2-m_i^2)(p^2-m_j^2)\ ,
\label{g}
\eea
and the loop integrals
\bea
\hspace{-0.5cm}\frac{i}{16\pi^2}\alpha_i &\equiv&
\int \frac{d^Dq}{(2\pi)^D}\frac{1}{(q^2-m_i^2)(q^2-\xi_i m_i^2)}\ , \\
\frac{i}{16\pi^2}\beta^{(0)}_{ij}(p^2) &\equiv&
\int \frac{d^4q}{(2\pi)^4}
\frac{1}{(q^2-m_i^2)(q^2-\xi_i m_i^2)[(q+p)^2-m_j^2]}\ , 
\eea
where $D=4-2\epsilon$.
\begin{figure}[t]
\vspace{1.cm}
\centerline{
\psfig{figure=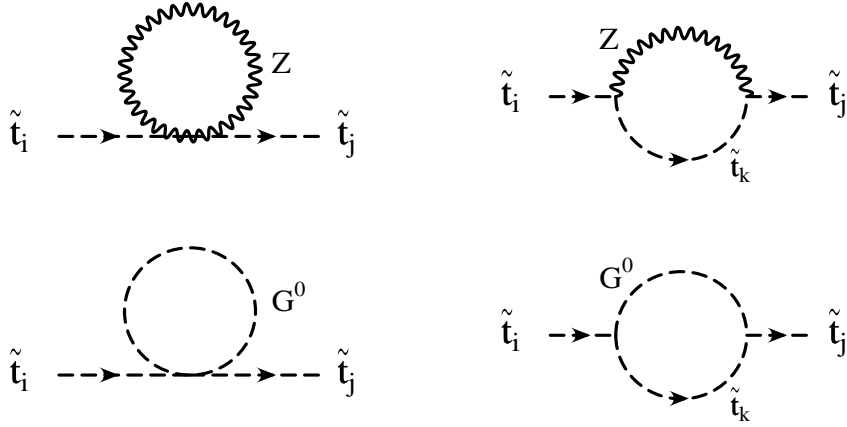,height=5cm,width=6cm,bbllx=8.cm,%
bblly=14.cm,bburx=15.cm,bbury=20.cm}}
\caption{
\noindent{\footnotesize
One-loop corrections to the top squark self-energies
that introduce a $\xi_Z$ dependence in $R_\xi$ gauges. 
Diagrams with $\xi_W$ dependence are quite similar
(with $Z\rightarrow W$, $G^0\rightarrow G^\pm$,
$\widetilde{t}_k\rightarrow\widetilde{b}_k$) and not shown.}}
\label{stopself}
\end{figure}

The Feynman diagrams that cause the gauge dependence of top squark
self-energies are depicted in Fig.~\ref{stopself}. Although this figure
shows only the diagrams that introduce a $\xi_Z$ dependence, the
$\xi_W$-dependent ones are quite similar.  We do not treat the
$\xi_{\gamma}$ and $\xi_g$ dependences since they are irrelevant for the
renormalization of the mixing angle. In addition, one should include the
gauge-dependent shifts of the vacuum expectation values (VEVs) $v_\alpha$ 
of the two Higgs bosons by tadpole graphs, for
the gauge-independent renormalization of the VEVs
\cite{Yamada,HiggsVEV,Freitas}.  The final result, given by
Eq.~(\ref{eq36}), satisfies the Nielsen identity (\ref{eq22}).  

The two
definitions of the scale-independent renormalized $\theta_{\tilt}$ given
in section 2 are gauge dependent, as can be shown by direct substitution
of Eq.~(\ref{eq36}) into Eq.~(\ref{ROS}) or Eq.~(\ref{Rp}). 
In more detail, for the on-shell scheme, although
\be
f_{ij}(m_i^2)+f_{ij}(m_j^2)=0\ ,
\label{ff}
\ee
one has 
\be
g_{ij}(m_i^2,m_k^2)\beta^{(0)}_{Z\tilt_k}(m_i^2)
+g_{ij}(m_j^2,m_k^2)\beta^{(0)}_{Z\tilt_k}(m_j^2)\neq
0\ ,
\label{noz}
\ee
(and a similar equation for the $W$ contribution);
while for the $p_*$ scheme, again
\be
f_{ij}(p_*^2)=0
\label{fp}
\ee
goes well, but
\be
g_{ij}(p_*^2)\beta^{(0)}_{Z\tilt_k}(p_*^2)
={1\over 4}(m_i^2-m_j^2)^2\beta^{(0)}_{Z\tilt_k}(p_*^2)\neq 0\
,
\ee
even if it gets ``closer'' than Eq.~(\ref{noz}) to being zero. 
(Of course, if $m_i^2=m_j^2$ the mixing angle is not well determined.)

\begin{figure}[t]
\vspace{1.cm}
\centerline{
\psfig{figure=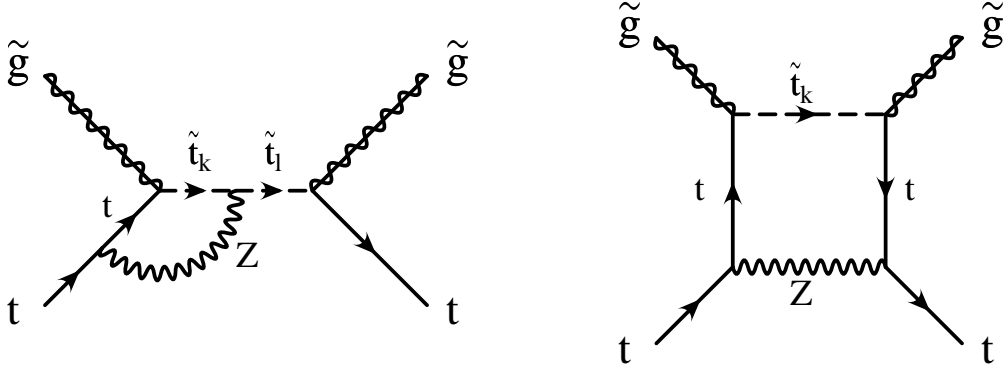,height=4cm,width=7cm,bbllx=8.cm,%
bblly=19.5cm,bburx=15.cm,bbury=23.5cm}}
\caption{
\noindent{\footnotesize
One-loop vertex and box corrections to the process
$\tilde{g}t\rightarrow\tilde{g}t$
that involve the $Z^0$ gauge boson and contribute to the pinched
top squark self-energies. Diagrams with $W^\pm$ corrections are quite
similar; simply replace $Z\rightarrow W$, $t\rightarrow b$,
$\widetilde{t}_k\rightarrow\widetilde{b}_k$ in internal lines. }}
\label{stopZ}
\end{figure}
\begin{figure}[t]
\vspace{1.cm}
\centerline{
\psfig{figure=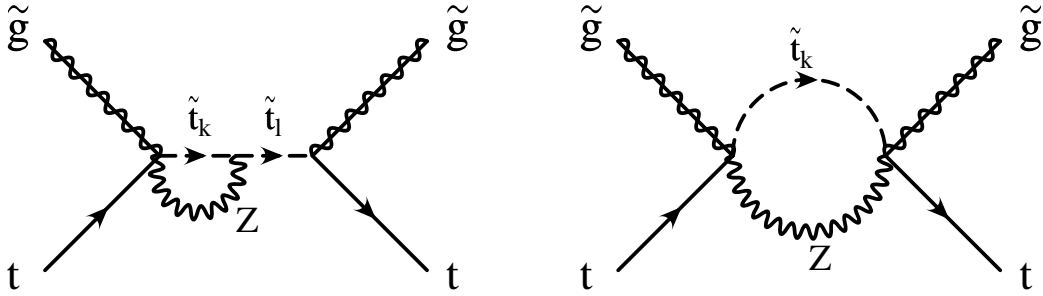,height=4cm,width=7cm,bbllx=8.cm,%
bblly=19.5cm,bburx=15.cm,bbury=23.5cm}}
\caption{
\noindent{\footnotesize
Diagrammatic representation of the pinched parts of the 
diagrams in fig.~\ref{stopZ}.}}
\label{stoppinch}
\end{figure}

When the top squark self-energies are embedded in the calculation of some
cross section at one loop, we know that their gauge dependence should be
cancelled by other one-loop contributions to the same process.
Processes with external on-shell top squarks were discussed in
\cite{Yamada}. Here we consider a more general 
case with off-shell top squarks. To avoid complications due to the mixing of 
external particles, we use the scattering amplitude of 
$t \tilde{g}\rightarrow t \tilde{g}$ with s-channel top squark exchange. 
In this particular case, the gauge dependence of the self-energy
corrections just discussed must be cancelled by that of the vertex
and box corrections (Fig.~\ref{stopZ}). 
The so-called pinch technique \cite{pinchsym,PaPi,Pa} 
extracts ``propagator-like'' contributions out of the vertex and 
box corrections. By triggering Ward-Takahashi identities at the 
gauge interaction vertices multiplied by longitudinal momenta, 
internal propagators in the loop are cancelled (or ``pinched''), giving
contributions that can be interpreted as shifts in the top squark
self-energies.
More precisely, in computing vertex corrections such as the one in
Fig.~\ref{stopZ} one uses identities like
\be
{1\over m_t -\qsla-\qsla_i}\qsla
=-1+{1\over m_t -\qsla-\qsla_i}(m_t-\qsla_i)\
, \label{xipinch}
\ee
where $q$ is the momentum of the virtual $Z$, and $q_i$ is the momentum of
the incoming top quark. The $\qsla$-independent operator $(m_t-\qsla_i)$,
moved to the right until it acts on the wave function of the external
top quark,
gives zero or something proportional to $m_t$ depending on the operators
that stand in its way. In the first term of Eq.~(\ref{xipinch}) the internal
top propagator has
been cancelled giving rise to a pinched contribution.
Figure~\ref{stoppinch} represents the pinched parts of the diagrams in 
Fig.~\ref{stopZ}, with the internal propagator of the top quark pinched.
The improved self-energies $\Pi_{ij}^P(p^2)$ that include these 
pinched terms are then gauge independent, do not modify the positions of
the poles and are process independent. 
\begin{figure}[t]
\vspace{1.cm}
\centerline{
\psfig{figure=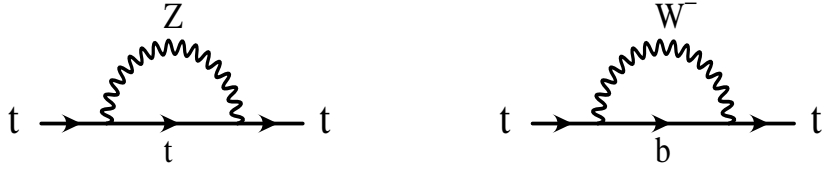,height=1.5cm,width=6cm,bbllx=8.cm,%
bblly=18.5cm,bburx=15.cm,bbury=20.cm}}
\caption{
\noindent{\footnotesize
One-loop wave-function-renormalization corrections for the top quark
that involve the $Z^0$ and $W^\pm$ gauge bosons.}}
\label{waveZW}
\end{figure}

Some pinch contributions come from the $(1-\xi_V)q_{\mu}q_{\nu}$
parts 
of the gauge boson propagators. The contribution from vertex 
corrections (including wave function corrections of external 
on-shell fermions, see Fig.~\ref{waveZW}) is 
\bea
\Delta\Pi^{\tilt}_{ij}(V) &=&
\frac{1}{16\pi^2}(2p^2-m_{\tilt_i}^2-m_{\tilt_j}^2)\nonumber
\\
&\times&\left\{ g_Z^2(1-\xi_Z)\chi^Z_{ik}\chi^Z_{jk}
\left[ \frac{1}{2}\alpha_Z - (p^2-m_{\tilt_k}^2) \beta^{(0)}_{Z\tilt_k}(p^2) 
\right] \right.\nonumber\\
&&\left. +2g_2^2(1-\xi_W) \chi^W_{ik}\chi^W_{jk}
\left[\frac{1}{2}\alpha_W -
(p^2-m_{\tilb_k}^2) \beta^{(0)}_{W\tilb_k}(p^2)  \right] \right\} \, .
\label{pstV}
\eea
In addition, the pinched box contribution is 
\bea
\Delta\Pi^{\tilt}_{ij}(B) &=&
\frac{1}{16\pi^2}(p^2-m_{\tilt_i}^2)(p^2-m_{\tilt_j}^2) \nonumber\\
&\times& \left[g_Z^2(1-\xi_Z)
\chi^Z_{ik}\chi^Z_{jk}\beta^{(0)}_{Z\tilt_k}(p^2) 
+2g_2^2(1-\xi_W)
\chi^W_{ik}\chi^W_{jk} \beta^{(0)}_{W\tilb_k}(p^2) \right] \, . 
\label{pstB}
\eea
Adding to the original self-energy (\ref{eq36}) the pinched contributions
(\ref{pstV}) and (\ref{pstB}), the gauge dependence exactly cancels and
the improved pinched self-energies, as well as the improved 
scale-independent mixing angle $\theta_{\tilt}$, are equal to
the simple $\xi=1$ results in the conventional calculation: 
\be
\Pi^{\tilt}_{ij}(p^2)+\Delta\Pi^{\tilt}_{ij}(V)+
\Delta\Pi^{\tilt}_{ij}(B)\equiv \Pi^{\tilt P}_{ij}(p^2)=
\left.\Pi^{\tilt}_{ij}(p^2)\right|_{\xi_Z=\xi_W=1}\ .
\ee
This is consistent with previous results 
\cite{Yamada} for general on-shell scalar particles (other than Higgs
bosons). 

However, there is another possible source of longitudinal momenta for 
pinching: the momentum-dependent $\tilt^*\tilt Z$ (and $\tilt\tilb W$)
couplings in Fig.~\ref{stopZ}. More explicitly, 
the factor $(2p+q)_{\mu}$ from the vertex $\tilt^*(-p-q)\tilt(p)Z_{\mu}(q)$ 
triggers the identity 
\be
{1\over m_t -\qsla-\qsla_i}(2\psla+\qsla)
=-1+{1\over m_t -\qsla-\qsla_i}[(m_t+\qsla_i)+2(\psla-\qsla_i)]\
, \label{xi1pinch}
\ee
which also gives a pinched contribution. [The $\qsla$-independent operators
$(m_t+\qsla_i)$ and $(\psla-\qsla_i)$ do not induce further pinching.] 
This type of additional pinch operation is known to be necessary to obtain 
gauge-independent self-energies for the Goldstone bosons in the Standard
Model \cite{Pa}. 
Notice that now this pinched contribution does not depend on
$\xi_{Z,W}$. Therefore, applying this additional pinching, the resulting 
improved self-energy is then different from the $\xi=1$ result by 
\bea
\label{Pinchedst}
\Delta_P\Pi^{\tilt}_{ij}(\tilt\tilt Z, \tilt\tilb W) &=& 
-\frac{1}{16\pi^2}(2p^2-m_{\tilt_i}^2-m_{\tilt_j}^2)\\
&\times& \left[ g_Z^2
\chi^Z_{ik}\chi^Z_{jk}B_0(p^2, m_Z^2, m_{\tilt_k}^2) 
+2g_2^2\chi^W_{ik}\chi^W_{jk}
B_0(p^2, m_W^2, m_{\tilb_k}^2) \right]\ ,
\nonumber
\eea
with
\be
\frac{i}{16\pi^2}B_0(p^2,m_i^2,m_j^2) =
\int \frac{d^Dq}{(2\pi)^D}
\frac{1}{(q^2-m_i^2)[(q+p)^2-m_j^2]}\ .
\ee
This might be regarded as a source of non-uniqueness of 
the pinch technique in theories with many scalar fields. 
The decomposition of the scalar-scalar-gauge vertices is necessary for 
the Goldstone boson self-energies in the Standard Model to be gauge 
independent and satisfy the naive Ward-Takahashi identities \cite{Pa}, but 
such arguments do not apply in the case of squarks (or other sfermions) 
and CP-even Higgs bosons.

It is easily seen that this additional term (\ref{Pinchedst}) 
does not affect the cancellation of the divergent correction 
to the mixing angle in the schemes in section~2 
[see Eqs.~(\ref{ff}) and (\ref{fp})]. Also, when applied to 
the gauge boson exchange contribution to the
$e^+e^-\rightarrow\tilt\tilt^*$ amplitudes, it is seen that the
decomposition of the $(\tilt\tilt Z, \tilt\tilb W)$ vertices does not
produce any additional contributions to be added to the gauge 
boson self-energies.  It is therefore not clear whether or not 
this type of pinching should be applied in this particular case.
We will go back to this problem in section 5.

We finally address the modification of the scale-independent mixing
angle by the pinch technique rearrangement. Substituting the forms of
the pinched corrections, Eqs.~(\ref{pstV}) and (\ref{pstB}), into the
counterterms in Eqs.~(\ref{ROS}) and (\ref{Rp}), it is amusing to note
that the $p_*$ scheme is not
affected by the vertex contribution, while the on-shell scheme is not
affected by the box contribution. 
An interesting point of the $p_*$
scheme is that it is insensitive to the arbitrariness in the treatment of
scalar-scalar-gauge vertices in the pinch operation. In particular, if 
the pinched term (\ref{Pinchedst}) is added to the top squark self-energies
then the $p_*$ definition of a scale- and gauge-independent mixing angle
for top squarks is not affected (with respect to the naive  $\xi=1$
result) while the on-shell definition should be modified to take into
account the effect of the contribution (\ref{Pinchedst}).

\section{Mixing angle of CP-even Higgs bosons}

The two CP-even Higgs scalars
$\eta_\alpha\equiv\sqrt{2}{\rm Re}H_\alpha^0-v_\alpha$ ($\alpha=1,2$) in
the MSSM mix with each other \cite{MSSM,GunionHaber} 
to form the mass eigenstates $h_i=(H^0, h^0)$.
They are related by a rotation matrix $R$ as
\be 
\eta_{\alpha}=R_{\alpha i}h_i = \left( \begin{array}{rr}  
\cos\alpha & -\sin\alpha \\ \sin\alpha & \cos\alpha \end{array} \right)
\left( \begin{array}{c} H^0 \\ h^0 \end{array} \right) .
\ee
The scale-independent renormalization of the mixing angle $\alpha$, along
the lines explained in section 2, inherits the gauge dependence of
the self-energy $\Pi^h_{ij}(p^2)$. In this section, we perform the 
improvement of $\Pi^h_{ij}(p^2)$ by the pinch technique. In contrast to
the case of top squarks,
the improved self-energies differ from the conventional
$\xi=1$ results, as is the case of the Higgs boson self-energy in the
Standard Model \cite{pinchHiggs}. We note in passing that the
gauge-independent Higgs boson self-energies are useful not only for a proper
definition of a scale-independent Higgs mixing angle $\alpha$, but also for
the treatment of s-channel resonant production of $(h^0,H^0)$ at 
photon \cite{gammagamma} or muon \cite{mumu} colliders.

Following the previous analysis in \cite{pinchHiggs}, we consider
$f \bar{f}\rightarrow f' \bar{f}'$ scatterings mediated by the
CP-even Higgs bosons. Working in the $R_\xi$ gauge, we obtain that the Higgs
boson self-energies depend on the gauge parameters through the following
expression
\bea
\Pi^h_{ij}(p^2)&=\left.\Pi^h_{ij}(p^2)\right|_{\xi_W=\xi_Z=1}&\nonumber\\
&+\displaystyle{{g_Z^2\over
64\pi^2}}(1-\xi_Z)&\hspace{-0.5cm}\left.\frac{}{}\right\{g_{ij}(p^2,m_A^2)
{\cal O}^{(1)}_{ij}\beta^{(0)}_{ZA}(p^2)
-f_{ij}(p^2)\delta_{ij}\alpha_Z
\nonumber\\
&&+\left.{1\over 2} g_{ij}(p^2,0){\cal O}^{(2)}_{ij}\left[
\beta^{(0)}_{ZZ}(p^2)+\beta^{(0)}_{Z,\xi Z}(p^2)\right]\right\}\nonumber\\
&+\displaystyle{{g_2^2\over 32\pi^2}}(1-\xi_W)
&\hspace{-0.5cm}\left.\frac{}{}\right\{g_{ij}(p^2,m_{H^\pm}^2)
{\cal O}^{(1)}_{ij}
\beta^{(0)}_{WH^\pm}(p^2)-
f_{ij}(p^2)\delta_{ij}\alpha_W\nonumber\\
&&+\left.{1\over 2} g_{ij}(p^2,0){\cal O}^{(2)}_{ij}\left[
\beta^{(0)}_{WW}(p^2)+\beta^{(0)}_{W,\xi W}(p^2)\right]\right\}\ ,
\label{hs}
\eea
where $f_{ij}$ and $g_{ij}$ have been defined in
Eqs.~(\ref{f},\ref{g}) and we have introduced the operators
\bea
{\cal O}_{ij}^{(1)}&\equiv &(R^h_{1i}\sin\beta-R^h_{2i}\cos\beta)
(R^h_{1j}\sin\beta-R^h_{2j}\cos\beta)\ ,\\ 
{\cal O}_{ij}^{(2)}& \equiv & (R^h_{1i}\cos\beta+R^h_{2i}\sin\beta)
(R^h_{1j}\cos\beta+R^h_{2j}\sin\beta)\ .
\eea
Notice in particular that these operators satisfy ${\cal
O}_{ij}^{(1)}+{\cal O}_{ij}^{(2)}=\delta_{ij}$ and 
${\cal O}_{ij}^{(1)}{\cal O}_{jk}^{(2)}=0$.

\begin{figure}[t]
\vspace{1.cm}
\centerline{
\psfig{figure=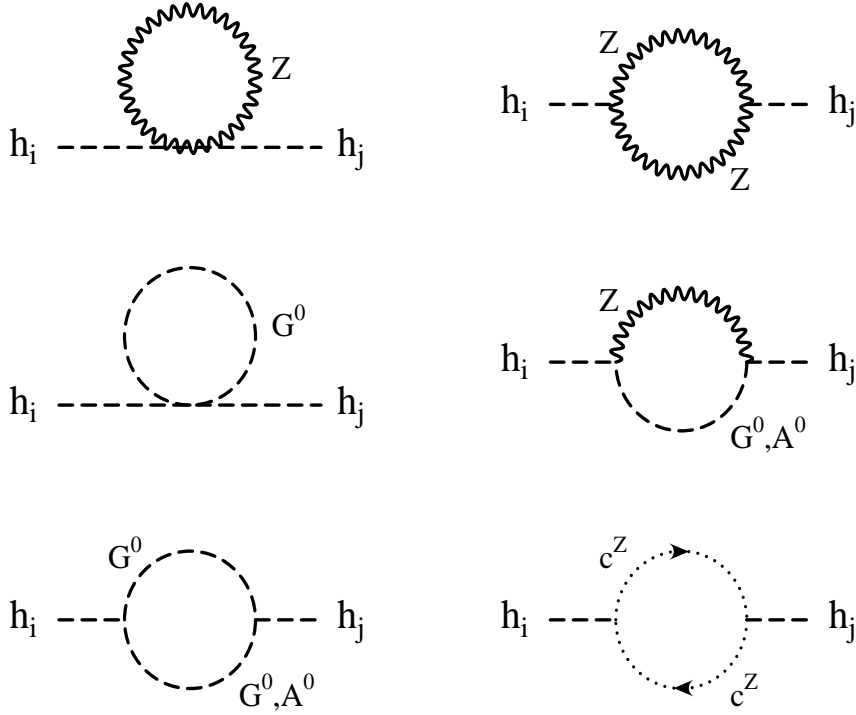,height=9cm,width=6cm,bbllx=8.cm,%
bblly=10.cm,bburx=15.cm,bbury=20.cm}}
\caption{
\noindent{\footnotesize
One-loop corrections to the Higgs boson self-energies
that introduce a $\xi_Z$ dependence in $R_\xi$ gauges. 
Diagrams with $\xi_W$ dependence are quite similar
(with $Z\rightarrow W$, $G^0\rightarrow G^\pm$, $A^0\rightarrow
H^\pm$ and $c^Z\rightarrow c^W$) and not shown.}}
\label{hself}
\end{figure}

The Feynman diagrams that introduce a dependence on $\xi_Z$ are presented
in Fig.~\ref{hself}. Those that give a $\xi_W$-dependent contribution
are quite similar and not shown. As in the case of top squarks in the
previous section, we also included the gauge-dependent shifts of the 
two Higgs boson VEVs by the corresponding tadpole graphs. In fact, the 
result (\ref{hs}) holds whenever the parameters ($m_Z$, $m_A$, $\tan\beta$) 
in the tree-level mass matrix of $\eta_\alpha$ are renormalized in 
gauge-independent ways.

\begin{figure}
\vspace{1.cm}
\centerline{
\psfig{figure=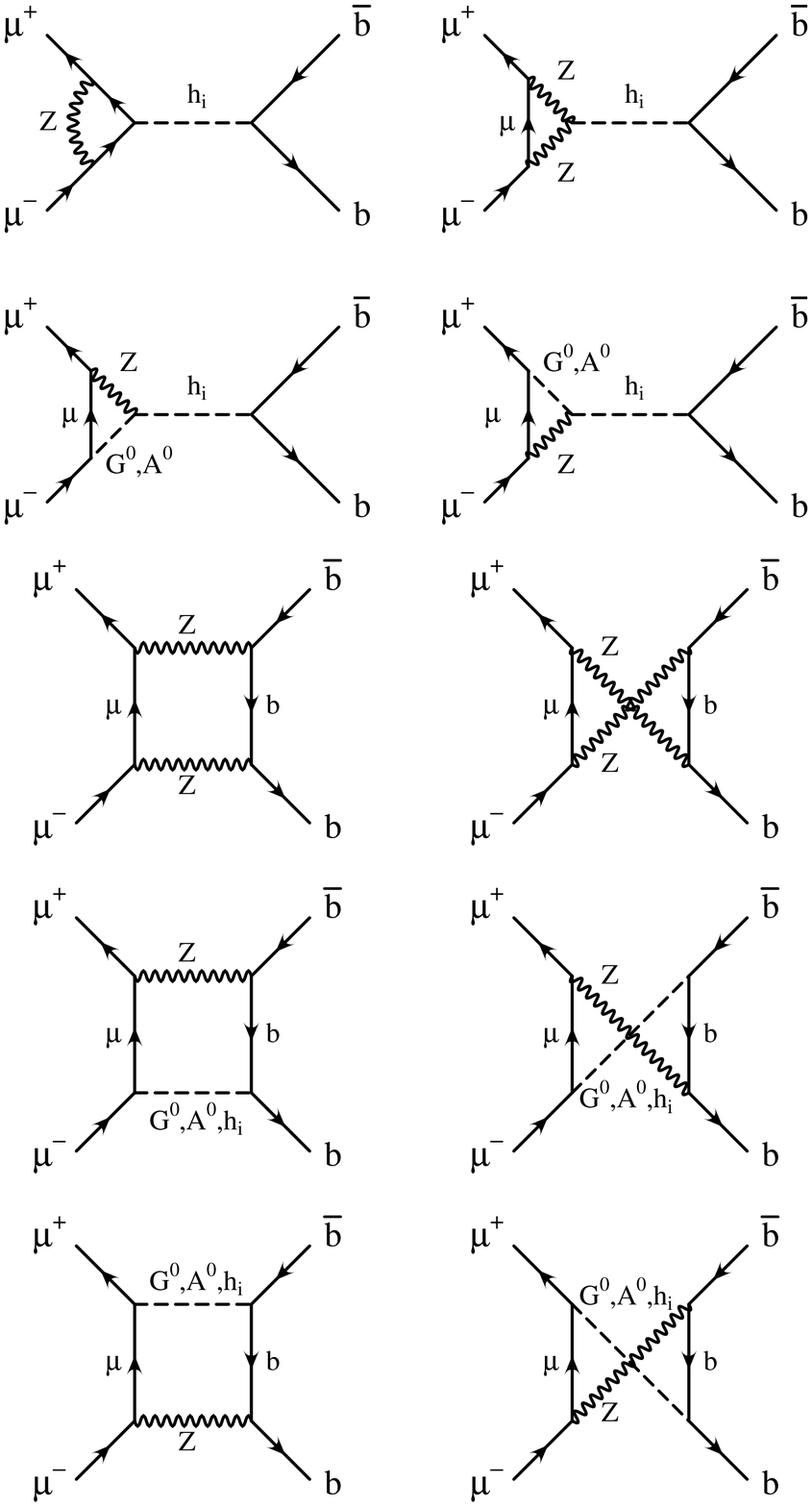,height=16cm,width=6cm,bbllx=8.cm,%
bblly=-3.cm,bburx=15.cm,bbury=20.cm}}
\caption{
\noindent{\footnotesize
One-loop vertex and box corrections to the process $\mu^+\mu^-\rightarrow
b\bar{b}$ that involve the $Z^0$ gauge boson and contribute to the pinched
Higgs boson self-energies. 
Diagrams with corrections to the $h_ib\bar{b}$ vertex are quite similar 
to those with $h_i\mu^+\mu^-$ vertex corrections and are not shown.}}
\label{Z}
\end{figure}
\begin{figure}[hbt]
\vspace{1.cm}
\centerline{
\psfig{figure=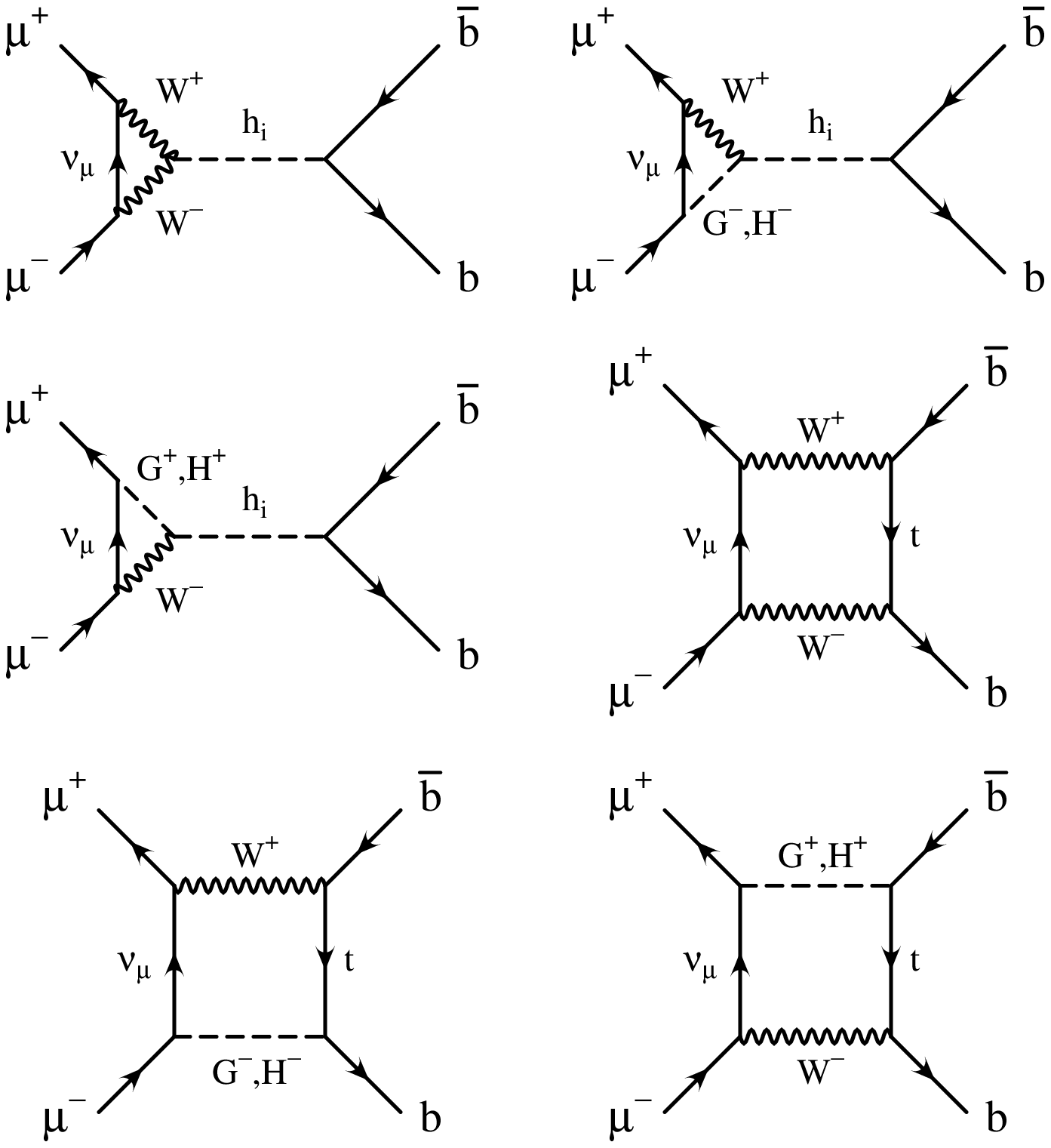,height=10cm,width=6cm,bbllx=8.cm,%
bblly=6.cm,bburx=15.cm,bbury=20.cm}}
\caption{
\noindent{\footnotesize
One-loop vertex and box corrections to the process $\mu^+\mu^-\rightarrow
b\bar{b}$ that involve the $W^\pm$ gauge bosons and contribute to the
pinched Higgs boson self-energies. 
Diagrams with corrections to the $h_ib\bar{b}$ vertex are quite similar 
to those with $h_i\mu^+\mu^-$ vertex corrections and are not shown.}}
\label{W}
\end{figure}

The ``propagator-like'' terms that correct the previous self-energies are
extracted from vertex and box corrections by the pinch technique. The
necessary diagrams are given in Figs.~\ref{Z} and \ref{W} for the
 case of $\mu^+\mu^-\rightarrow b\bar{b}$ scattering.  
Note that one also has to analyze processes with external up-type
($I_3=+1/2$) fermions for the correct assignment of pinch terms to
$\Pi^h_{11}$, $\Pi^h_{22}$, and $\Pi^h_{12}$. Box diagrams with
$(Z,h_i)$, as
well as parts of the box diagrams with $W^{\pm}$, give propagator-like
terms with pseudoscalar couplings to fermions. These are pinch terms for
self-energies of gauge and $(A^0,G^0)$ bosons, and not relevant for our
study. In addition to the diagrams shown, one should also include the
pinching coming from wave function renormalization of external legs.

As in the previous section, there are two sources of momenta for 
pinching:
 the $(1-\xi_V)q_{\mu}q_{\nu}$ parts of the gauge boson propagators and
gauge-scalar-scalar vertices. The former parts generate the following
pinch terms from vertex corrections
(including fermion wave function corrections):
\bea
\Delta\Pi^h_{ij}(V)&=& -{1\over 128
\pi^2}(2p^2-m_i^2-m_j^2)\nonumber\\
&\times&
\left[\frac{}{}g_Z^2(1-\xi_Z)\left\{2(p^2-m_A^2){\cal O}^{(1)}_{ij}
\beta^{(0)}_{ZA}(p^2) -\delta_{ij}\alpha_Z\right.\right.\nonumber\\
&&+\left.{\cal O}^{(2)}_{ij}\left[
(p^2+2m_Z^2)\beta^{(0)}_{ZZ}(p^2)+p^2\beta^{(0)}_{Z,\xi Z}(p^2)\right]
\right\}\nonumber\\
&&+2g_2^2(1-\xi_W)
\left\{2(p^2-m_{H^\pm}^2){\cal O}^{(1)}_{ij}
\beta^{(0)}_{WH^\pm}(p^2) -\delta_{ij}\alpha_W \right.\nonumber\\
&&\left.\left.+{\cal O}^{(2)}_{ij}\left[
(p^2+2m_W^2)\beta^{(0)}_{WW}(p^2)+p^2\beta^{(0)}_{W,\xi W}(p^2)\right]   
\right\}\frac{}{}\right]\ ,
\label{phV}
\eea
and from box correction diagrams
\bea
\Delta\Pi^h_{ij}(B)&=& {1\over 64\pi^2}(p^2-m_i^2)(p^2-m_j^2) \nonumber\\
&\times & \left[ g_Z^2(1-\xi_Z)
\left\{ {\cal O}^{(1)}_{ij}\beta^{(0)}_{ZA}(p^2)
+{1\over 2} {\cal O}^{(2)}_{ij}\left[
\beta^{(0)}_{ZZ}(p^2)+\beta^{(0)}_{Z,\xi Z}(p^2)\right]\right\} \right.
\label{phB}
\\
&& + \left. 2g_2^2(1-\xi_W)
\left\{ {\cal O}^{(1)}_{ij}\beta^{(0)}_{WH^\pm}(p^2)
+{1\over 2} {\cal O}^{(2)}_{ij}\left[
\beta^{(0)}_{WW}(p^2)+\beta^{(0)}_{W,\xi W}(p^2)\right] \right\} \right]
\, .
\nonumber
\eea

Comparing to Eq.~(\ref{hs}) one sees that the above two contributions
are not enough to cancel the $\xi$ dependence of $\Pi^h_{ij}$. In
fact, pinching by the momenta in the $G^0$-$Z$-$h_i$ and
$G^{\pm}$-$W^{\mp}$-$h_i$ vertices in the vertex corrections of
Figs.~\ref{Z} and \ref{W}, is necessary to obtain gauge-independent
self-energies, as in previous studies of the
self-energies for massive
gauge bosons and their Golstone modes \cite{Pa}, as well as the Standard
Model Higgs boson \cite{pinchHiggs}. This type of pinching generates
additional contributions to the Higgs boson self-energies:
\bea
\Delta\Pi^h_{ij}(hZG^0,hW^{\pm}G^{\mp})&=& -{1\over
64\pi^2}(2p^2-m_i^2-m_j^2){\cal O}^{(2)}_{ij}\nonumber\\
& \times & \left[ g_Z^2
B_0(p^2, m_Z^2, \xi_Z m_Z^2)+ 2g_2^2
B_0(p^2,m_W^2,\xi_W m_W^2) \right] \, .
\eea
As a result, the modified self-energies become $\xi$ independent, but
different from the $\xi=1$ form (unlike what happened for top squarks in
section~3) by
\bea
\Delta_P\Pi^h_{ij}(p^2) &=& 
- {1\over 64\pi^2}(2p^2-m_i^2-m_j^2){\cal O}^{(2)}_{ij} \nonumber\\ 
&\times& \left[g_Z^2 B_0(p^2, m_Z^2, m_Z^2)
+ 2g_2^2 B_0(p^2,m_W^2,m_W^2) \right] .
\label{hpinch}
\eea
Although all the gauge dependence is now cancelled, the improved 
self-energy  has one unsatisfactory property. 
The anomalous dimensions of $h_i$ determined
by the pinched self-energy $\Pi^h_{ij}+\Delta_P\Pi^h_{ij}$ are not
diagonal 
in the gauge eigenbasis $\eta_{\alpha}$, and this  might cause problems in
renormalization.

We now consider the last possible source for pinching:
the $A^0$-$Z$-$h_i$ and $H^{\pm}$-$W^{\mp}$-$h_i$
vertices in vertex corrections (Figs.~\ref{Z} and \ref{W}). As in the
case of top squarks, there is the freedom of whether or not
to perform the pinching by these momenta. Since the additional pinch
terms are gauge independent, it looks unnecessary to include
these contributions. However, if one includes them, the resulting
self-energies are
\bea
\Pi^{h P}_{ij}(p^2) &\equiv&
\Pi^{h}_{ij}(p^2)+\Delta \Pi^{h}_{ij}(V,B)
+\Delta\Pi^h_{ij}(hZG^0,hW^{\pm}G^{\mp})
+\Delta\Pi^h_{ij}(hZA^0,hW^{\pm}H^{\mp})
\nonumber\\
&=& \Pi^h_{ij}(p^2)_{\xi=1}
- {1\over 64 \pi^2}(2p^2-m_i^2-m_j^2) \nonumber\\
& \times& \left[ g_Z^2 \left\{
{\cal O}^{(2)}_{ij} B_0(p^2, m_Z^2, m_Z^2) +
{\cal O}^{(1)}_{ij}B_0(p^2, m_Z^2, m_A^2) \right\} \right. \nonumber\\
&& \left. + 2g_2^2 \left\{
{\cal O}^{(2)}_{ij}B_0(p^2,m_W^2,m_W^2) +
{\cal O}^{(1)}_{ij}B_0(p^2,m_W^2,m_{H^{\pm}}^2) \right\} \right] 
\label{Pinchedh}
\eea
and the anomalous dimensions then become diagonal in the
gauge basis. Moreover, these modified anomalous dimensions
match with the running of the Higgs boson VEVs $v_\alpha$
in the $\xi=1$ gauge:
the discrepancy between the running of $\eta_\alpha$ and of $v_\alpha$
\cite{deltav,CPR} is cancelled by the additional term in
Eq.~(\ref{Pinchedh}). We therefore conclude that
Eq.~(\ref{Pinchedh}) gives the form of the gauge-independent
self-energies that should be preferred. In the next section, these nice
features of the self-energies (\ref{Pinchedh}) are
explained in the framework of the background field method.

It is quite reasonable to adopt the well motivated definition of the Higgs
boson self-energies given by Eq.~(\ref{Pinchedh}) to compute scale-independent
Higgs mixing angles that are also gauge independent (as explained in
section~2). As was discussed at the end of the previous section for the
case of the mixing angle between top squarks, in the $p_*$ scheme this
modification of the self-energies does not affect the definition of the
mixing angle (it is the same as computed from the non-pinched two-point
function in the $\xi=1$ gauge). In contrast, the scale-independent
definition in the on-shell scheme gets additional corrections from the new
terms in Eq.~(\ref{Pinchedh}).

\section{Relation to the background field method}

It has been shown \cite{Hashimoto,Denner} that the self-energies of 
gauge bosons improved by the pinch technique can be obtained 
as a special $\xi=1$ case of the background field 
method \cite{BFM}. This relation also holds for the self-energy of 
the standard model Higgs boson \cite{pinchHiggs}. In this section, 
we consider the background field method in theories with many scalar 
fields, such as the MSSM, and show that the freedom in the treatment
of 
the scalar-scalar-gauge vertices in the pinch technique, discussed 
in previous sections, corresponds to freedom in the 
choice of the gauge fixing function. 

In the background field method, one first splits the 
gauge bosons $V_{\mu}^a$ ($a$ is the gauge group index of the 
adjoint representation) and scalar bosons $\Phi_I$ ($I$ 
denotes one gauge multiplet) into background (with hat) and 
quantum (without hat) fields, as 
\be
V_{\mu}^a \rightarrow \widehat{V}_{\mu}^a + V_{\mu}^a, \;\;\;
\Phi_I \rightarrow \widehat{\Phi}_I + \Phi_I. 
\ee
The splitting of fermions is trivial and not shown here. When the scalar 
bosons have nonvanishing VEVs, these are assigned to the 
background fields $\widehat{\Phi}_I$. (In other words, quantum fields have 
no VEVs.)

The essential point of the background field method is a clever 
choice of the gauge fixing function $F^a$, 
\be
F^a = \partial^{\mu} V^a_{\mu} - i g [\widehat{V}^{\mu}, V_{\mu}]^a 
- i g \xi_V \sum_I ( \widehat{\Phi}^{\dagger}_I T^{(I)a} \Phi_I 
- \Phi^{\dagger}_I T^{(I)a} \widehat{\Phi}_I ) , \label{BFMgauge}
\ee
where $T^{(I)a}$ is the group generator for $\Phi_I$. 
This is a natural extension of the usual $R_{\xi}$ gauge fixing and 
manifestly keeps local gauge invariance for background fields. 
Note that we have to set $\xi_Z=\xi_W=\xi_{\gamma}$ to preserve the 
SU(2)$\times$U(1) gauge symmetry. 

The one-particle-irreducible vertex functions, given by the loops 
with external background fields, then satisfy the naive, tree-level-like  
Ward-Takahashi identities \cite{Hashimoto,Denner,BFM}. 
In the standard model, the background field method with $\xi=1$ gives 
the same one-loop self-energies and vertex functions as the pinch 
technique \cite{pinchHiggs,Hashimoto,Denner}. This result can be 
understood by observing that 
the change of the propagators and three-point functions, in going from 
the conventional $R_{\xi}$ to the gauge in Eq.~(\ref{BFMgauge}),
corresponds 
to the splitting of longitudinal momenta for the pinching. 

In applying the gauge fixing (\ref{BFMgauge}) to theories with many scalar 
fields, like the MSSM, there is one novel source of arbitrariness 
which is irrelevant for the standard model. The gauge 
fixing function $F^a$ should include the scalar fields which have 
gauge-symmetry-breaking VEVs, exactly as in Eq.~(\ref{BFMgauge}), to cancel 
the mixing of gauge and scalar bosons. However, one may also include 
scalar fields with no VEVs into Eq.~(\ref{BFMgauge}). Unlike what happens in 
the conventional $R_{\xi}$ gauge, this inclusion causes a nontrivial 
change of the Feynman rules in the scalar-scalar-gauge vertices. If a 
scalar field $\phi$ is not included in $F^a$, the vertex 
$\phi^*(-p-q)\widehat{\phi}(p)V_{\mu}(q)$ is proportional to $(2p+q)_{\mu}$. 
On the other hand, if $\phi$ is included in $F^a$, the same vertex becomes 
proportional to $2p_{\mu}$. Compared to the calculation in sections 3 and 4, 
it is clearly seen that this modification just amounts to the additional 
pinching by the momentum from the $\phi^*\phi V_{\mu}$ gauge vertex. 
In the rearrangement of loop corrections by the pinch technique, 
consequently, the momentum in the $\phi^* \phi V_{\mu}$ gauge vertex, 
with one $\phi$ not carrying loop momentum, should be used 
for pinching if and only if $\phi$ is included in the gauge fixing 
function $F^a$. 

In the MSSM, all sources of gauge symmetry breaking can be 
put together into one ``standard-like'' scalar doublet, 
\be
H_{\rm SM}\equiv 
\cos\beta ( -H_1^+, H_1^{0*}) + \sin\beta ( H_2^+, H_2^0), 
\ee
which contains a combination 
$\phi_{SM}=\cos(\alpha-\beta)H^0-\sin(\alpha-\beta)h^0$, the Goldstone 
modes $G^+$, $G^0$ and the VEV $v=\sqrt{v_1^2+v_2^2}$:
\be
H_{\rm SM}=[G^+, (\phi_{SM}+v+i G^0)/\sqrt{2}]\ .
\ee 
The minimal choice for the background gauge fixing is to include 
only the above $H_{\rm SM}$ into Eq.~(\ref{BFMgauge}). 
This choice reproduces the first version [Eq.~(\ref{hpinch})] of the 
pinched self-energies for Higgs bosons. 
In this choice, the gauge fixing (\ref{BFMgauge}) 
has hard breaking of a discrete symmetry under 
$(H_1,H_2)\rightarrow(-H_1,H_2)$, causing non-diagonal 
anomalous dimensions in Eq.~(\ref{hpinch}). Alternatively, we may include both 
$H_1$ and $H_2$  in $F^a$ with equal weight, as 
$(\widehat{H}_1^{\dagger}T^aH_1+\widehat{H}_2^{\dagger}T^aH_2-{\rm h.c.})$. 
This choice reproduces the theoretically preferable result
in Eq.~(\ref{Pinchedh}). 
Since this gauge fixing preserves the discrete symmetry shown above, the 
improved anomalous dimensions given by (\ref{Pinchedh}) should be 
diagonal in the gauge basis. Moreover, unlike what happens in the
conventional 
$R_{\xi}$ gauge \cite{deltav}, there are no sources of
discrepancy between the 
running of $\eta_\alpha$ and that of $v_\alpha$. We therefore conclude that, 
in using the pinch technique for the MSSM, we should 
decompose all Higgs-Higgs-gauge vertices for the pinching. 

We finally comment on the same kind of freedom in
the top squark self-energies discussed in section 3, due to
the possible additional pinching  by momenta 
in the $\tilt^*\tilt Z$ and $\tilt^*\tilb W$ couplings
[Eq.~(\ref{Pinchedst})]. 
It is now evident that this freedom exactly corresponds to 
that of including or not including squarks into $F^a$ in the background field
method. 
In contrast with the case of the Higgs bosons just discussed, we could 
find no theoretical arguments for the inclusion of squarks in the gauge
fixing function. 
Therefore, for simplicity, we prefer not to use for pinching the momenta
in the scalar-scalar-gauge couplings, except for the Higgs bosons.

\section{Summary and conclusions}

In this paper we have shown how to improve existing definitions of mixing
angles for scalar particles that are renormalization-scale and
process independent to make them also gauge independent. This we achieve
by applying the pinch technique in order to obtain gauge-independent
self-energies from which gauge-independent mixing angles are obtained. Further
assistance from the background field method was required to sort out some
arbitrariness (that the pinch technique by
itself could not resolve) concerning what pinched corrections to include.

We applied this procedure to two particular cases in the framework of the
Minimal Supersymmetric Standard Model where such improvement will be
relevant: the mixing of top squarks and of Higgs bosons. From our one-loop
calculation of the pinched contributions to the self-energies in the $R_\xi$
gauge we conclude the following.
 
For the top squark sector, we advocate a simple self-energy improved
by pinching 
that exactly coincides with the unpinched self-energy evaluated in 
the Feynman gauge. 
The prescription for a scale-independent top squark mixing angle that is also 
gauge-independent is therefore to use either the 
on-shell \cite{GuaschEberl} or the $p_*$ scheme \cite{EN} 
with the usual (unpinched) self-energies in the Feynman gauge.

For the Higgs boson sector, the self-energy improved by pinching
does not coincide 
with the unpinched self-energy in the Feynman gauge but contains additional 
pinched terms.  In the on-shell scheme, these new terms modify the 
counterterm (\ref{ROS}) of the mixing angle from the form
in the Feynman gauge.
In contrast, these new terms vanish in the $p_*$ scheme and 
therefore in that scheme the gauge-independent definition of 
the Higgs boson mixing angle is again obtained from the unpinched 
self-energies in the Feynman gauge. 
\vspace{0.5cm}

{\bf Acknowledgments:} J.R.E. thanks Ignacio Navarro for participating in the
early stages of this work, Alberto Casas for a careful reading of
the manuscript
and Fantina Madricardo for useful correspondence.
Y.Y. was supported in part by the Grant-in-aid for 
Scientific Research from the Ministry of Education, Culture, Sports, 
Science, and Technology of Japan, No.~14740144.

\newpage


\begin{thebibliography}{99}
\bibitem{MSSM}
H.~P.~Nilles,
Phys.\ Rept.\  {\bf 110} (1984) 1; \\
H.~E.~Haber and G.~L.~Kane,
Phys.\ Rept.\  {\bf 117} (1985) 75;\\
S.~P.~Martin,
[hep-ph/9709356].
\bibitem{GuaschEberl} 
J.~Guasch, J.~Sol\`a and W.~Hollik,
Phys.\ Lett.\ B {\bf 437} (1998) 88
[hep-ph/9802329];\\
H.~Eberl, S.~Kraml and W.~Majerotto,
JHEP {\bf 9905} (1999) 016
[hep-ph/9903413];\\
J.~Guasch, W.~Hollik and J.~Sol\`a,
Phys.\ Lett.\ B {\bf 510} (2001) 211
[hep-ph/0101086].
\bibitem{EN}
J.~R.~Espinosa and I.~Navarro,
Phys.\ Rev.\ D {\bf 66} (2002) 016004
[hep-ph/0109126].
\bibitem{Yamada}
Y.~Yamada,
Phys.\ Rev.\ D {\bf 64} (2001) 036008
[hep-ph/0103046].
\bibitem{CKM}
P.~Gambino, P.~A.~Grassi and F.~Madricardo,
Phys.\ Lett.\ B {\bf 454} (1999) 98
[hep-ph/9811470];\\
B.~A.~Kniehl, F.~Madricardo and M.~Steinhauser,
Phys.\ Rev.\ D {\bf 62} (2000) 073010
[hep-ph/0005060];\\
A.~Barroso, L.~Br\"ucher and R.~Santos,
Phys.\ Rev.\ D {\bf 62} (2000) 096003
[hep-ph/0004136].
\bibitem{Pila}
A.~Pilaftsis,
Phys.\ Rev.\ D {\bf 65} (2002) 115013 [hep-ph/0203210].
\bibitem{pinchsym}
J.~M.~Cornwall, in {\it Proceedings of the French-American Seminar on
Theoretical Aspects of Quantum Chromodynamics}, Marseille, France, 1981,
edited by J.W. Dash (Centre de Physique Th\'eorique, Marseille, 1982);
Phys.\ Rev.\ D {\bf 26} (1982) 1453;\\
J.~M.~Cornwall and J.~Papavassiliou,
Phys.\ Rev.\ D {\bf 40} (1989) 3474.
\bibitem{PaPi}
J.~Papavassiliou,
Phys.\ Rev.\ D {\bf 41} (1990) 3179;\\
G.~Degrassi and A.~Sirlin,
Phys.\ Rev.\ D {\bf 46} (1992) 3104;\\
G.~Degrassi, B.~A.~Kniehl and A.~Sirlin,
Phys.\ Rev.\ D {\bf 48} (1993) 3963;\\
J.~Papavassiliou and A.~Pilaftsis,
Phys.\ Rev.\ Lett.\  {\bf 75} (1995) 3060
[hep-ph/9506417];
Phys.\ Rev.\ D {\bf 53} (1996) 2128
[hep-ph/9507246];
Phys.\ Rev.\ D {\bf 54} (1996) 5315
[hep-ph/9605385];\\
D.~Binosi and J.~Papavassiliou,
Phys.\ Rev.\ D {\bf 66} (2002) 025024
[hep-ph/0204128];
Phys.\ Rev.\ D {\bf 66} (2002) 076010
[hep-ph/0204308].
\bibitem{Pa}
J.~Papavassiliou,
Phys.\ Rev.\ D {\bf 50} (1994) 5958
[hep-ph/9406258].
\bibitem{nielsen}
N.~K.~Nielsen,
Nucl.\ Phys.\ B {\bf 101} (1975) 173;\\
O.~Piguet and K.~Sibold,
Nucl.\ Phys.\ B {\bf 253} (1985) 517.
\bibitem{Gambino}
P.~Gambino and P.~A.~Grassi,
Phys.\ Rev.\ D {\bf 62} (2000) 076002.
[hep-ph/9907254].
\bibitem{pierce}
D.~M.~Pierce, J.~A.~Bagger, K.~T.~Matchev and R.-J.~Zhang,
Nucl.\ Phys.\ B {\bf 491} (1997) 3
[hep-ph/9606211].
\bibitem{HiggsVEV}
G.~Degrassi and A.~Sirlin,
Nucl.\ Phys.\ B {\bf 383} (1992) 73;\\
R.~Hempfling and B.~A.~Kniehl,
Phys.\ Rev.\ D {\bf 51} (1995) 1386,
[hep-ph/9408313].
\bibitem{Freitas}
A.~Freitas and D.~St\"ockinger,
Phys.\ Rev.\ D {\bf 66} (2002) 095014
[hep-ph/0205281].
\bibitem{pinchHiggs}
J.~Papavassiliou and A.~Pilaftsis,
Phys.\ Rev.\ D {\bf 58} (1998) 053002
[hep-ph/9710426].
\bibitem{GunionHaber}
J.~F.~Gunion and H.~E.~Haber,
Nucl.\ Phys.\ B {\bf 272} (1986) 1 
[Erratum-ibid.\ B {\bf 402} (1993) 567].
\bibitem{gammagamma}
D.~L.~Borden, D.~A.~Bauer and D.~O.~Caldwell,
Phys.\ Rev.\ D {\bf 48} (1993) 4018; \\
J.~F.~Gunion and H.~E.~Haber,
Phys.\ Rev.\ D {\bf 48} (1993) 5109.
\bibitem{mumu}
V.~D.~Barger, M.~S.~Berger, J.~F.~Gunion and T.~Han,
Phys.\ Rev.\ Lett.\  {\bf 75} (1995) 1462
[hep-ph/9504330]; 
Phys.\ Rept.\  {\bf 286} (1997) 1
[hep-ph/9602415];
Phys.\ Rev.\ Lett.\  {\bf 78} (1997) 3991
[hep-ph/9612279];\\
A.~Pilaftsis,
Phys.\ Rev.\ Lett.\  {\bf 77} (1996) 4996
[hep-ph/9603328].
\bibitem{deltav}
M.~Okawa,
Prog.\ Theor.\ Phys.\  {\bf 60} (1978) 1175;\\
A.~Schilling and P.~van Nieuwenhuizen,
Phys.\ Rev.\ D {\bf 50} (1994) 967 [hep-th/9408125].
\bibitem{CPR}
P.~H.~Chankowski, S.~Pokorski and J.~Rosiek,
Phys.\ Lett.\ B {\bf 274} (1992) 191;
Nucl.\ Phys.\ B {\bf 423} (1994) 437 [hep-ph/9303309].
\bibitem{Hashimoto}
S.~Hashimoto, J.~Kodaira, Y.~Yasui and K.~Sasaki,
Phys.\ Rev.\ D {\bf 50} (1994) 7066
[hep-ph/9406271].
\bibitem{Denner}
A.~Denner, G.~Weiglein and S.~Dittmaier,
Phys.\ Lett.\ B {\bf 333} (1994) 420
[hep-ph/9406204]; 
Nucl.\ Phys.\ B {\bf 440} (1995) 95 [hep-ph/9410338].
\bibitem{BFM}
B.~S.~Dewitt,
Phys.\ Rev.\  {\bf 162} (1967) 1195; \\
G. 't Hooft, Acta Universitatis Wratislavensis {\bf 368} (1976) 345;\\ 
H.~Kluberg-Stern and J.~B.~Zuber,
Phys.\ Rev.\ D {\bf 12} (1975) 482;
Phys.\ Rev.\ D {\bf 12} (1975) 3159; \\ 
L.~F.~Abbott,
Nucl.\ Phys.\ B {\bf 185} (1981) 189; 
Acta Phys.\ Polon.\ B {\bf 13} (1982) 33.
\end{thebibliography}
\end{document}